\documentstyle[mprocl]{article}

\def\be{\begin{equation}}
\def\ee{\end{equation}}
\def\bea{\begin{eqnarray}}
\def\eea{\end{eqnarray}}

\begin{document}
\title{THE RELIABILITY HORIZON}
\author{MATT VISSER}
\address{Physics Department, Washington University, \\
Saint Louis, Missouri 63130--4899, USA}

\maketitle
\abstracts{
The {\em ``reliability horizon''} for semi-classical quantum gravity
quantifies the extent to which we should trust semi-classical quantum
gravity, and gives a handle on just where the {\em ``Planck regime''}
resides. The key obstruction to pushing semi-classical quantum gravity
into the Planck regime is often the existence of large metric
fluctuations, rather than a large back-reaction. 
}

\section{Introduction}

Semi-classical quantum gravity is the approximation wherein we keep
the gravitational field classical, while quantizing everything
else. Where should we stop believing this approximation?
Qualitatively, the answer to this question has been known since the
pioneering work of Wheeler.~\cite{Wheeler55,Wheeler57} We should
certainly stop believing semi-classical quantum gravity once we enter
the Planck regime.  The subtleties arise in recognizing the onset of
Planck scale physics.

To quantify these issues I have recently introduced the concepts of
the {\em ``reliable region''}, the {\em ``reliability boundary''}, and
the {\em ``reliability horizon''}.~\cite{Reliable} These concepts are
defined in a way that is similar to standard concepts of classical
general relativity: the chronology-violating region; chronology
boundary; and chronology horizon respectively.  Physically, the
reliability horizon is characterized by the onset of {\em either}
large metric fluctuations (Planck scale curvature fluctuations) {\em
or} a large back-reaction (Planck scale expectation value for the
curvature). In many situations the onset of large metric fluctuations
precedes the onset of large back-reaction.~\cite{Reliable,Roman-ring}

\section{The chronology horizon}

Let $\gamma$ be any geodesic (spacelike, null, or timelike) that
connects some point $x$ to itself. Let $\sigma_\gamma(x,y)$ denote
the relativistic interval from $x$ to $y$ along the geodesic $\gamma$.

\begin{equation}
\sigma_\gamma(x,y) = 
\left\{
\matrix{+s^2               &    \hbox{if the geodesic is spacelike,}\cr
        \hphantom{+}0      &    \hbox{if the geodesic is lightlike,}\cr
        -\tau^2            &    \hbox{if the geodesic is timelike.} \cr}
\right.
\end{equation}
Define level sets $\Omega(\ell^2)$ by

\begin{equation}
\Omega(\ell^2) \equiv 
\left\{x:\exists \gamma\neq0 | \sigma_\gamma(x,x) \leq \ell^2\right\}.
\end{equation}
---The set $\Omega(0)$ is called the {\em ``chronology violating
region''}.\\ 
---The set ${\cal B} \equiv \partial[\Omega(0)]$ is the {\em ``chronology
boundary''}. This is the boundary that we will have to cross in
order to actively participate in time travel effects.\\
---The set ${\cal H}^+ \equiv \partial [J^+(\Omega(0))]$ is the {\em
``chronology horizon''}. It is the boundary of the future of the
chronology violating region.  This is the boundary that we will
have to cross in order to passively participate in time travel
effects. 

These definitions only make sense if we are dealing with a fixed
uquantized Lorentzian geometry---this is exactly the statement that
we are dealing with semi-classical quantum  gravity.

\section{The reliability horizon}

I define the reliability horizon in three stages:~\cite{Reliable}

{\bf Definition 1a:} Using notation as above, let ${\cal U} \equiv
\Omega(+\ell_{Planck}^2)$ be the {\em ``unreliable region''}. It
consists of those points $x$ that are connected to themselves by
spacelike geodesics as short as, or shorter than, one Planck length.

The entire thrust of this definition is that it gives an invariant and
unambiguous meaning to the notion ``within a Planck length of the
chronology horizon'', an invariant interpretation of this phrase being
necessary before it is possible to decide where the Planck regime
resides.

{\bf Definition 1b:} Let ${\cal B}_{Planck} \equiv \partial [{\cal
U}] \equiv \partial [\Omega(+\ell_{Planck}^2)]$ be the {\em
``reliability boundary''}---this is the boundary that we will have
to cross in order to actively probe the unreliable region.

{\bf Definition 1c:} The set ${\cal H}^+_{Planck} \equiv
\partial[J^+({\cal U})] \equiv \partial [
J^+(\Omega(+\ell_{Planck}^2))]$ is the {\em ``reliability
horizon''}---it is the boundary of the future of the unreliable
region.  This is the boundary that we will have to cross in order to
passively probe the unreliable region.

{\bf Justification:} At this stage these are merely definitions, they
do not carry any weight until we physically justify the
terminology.~\cite{Reliable} If we take Einstein gravity and linearize
it about the background we are interested in, we can then ask how the
linearized gravitons behave as quantum fields on this background
geometry.  The resulting quantum field theory is well known to be
non-renormalizable with a dimensionful coupling constant given by the
Planck mass.~\cite{Wheeler55}$^{\!,\,}$\cite{Wheeler57} Once we enter
the unreliable region these linearized gravitons are subject to Planck
scale physics which in this non-renormalizable theory is definitely a
disaster.  Inside the unreliable region the linearized gravitons will
be strongly interacting (and also unitarity violating) and will
thereby lead to Planck scale fluctuations in the
curvature,~\cite{Reliable} even if the expectation value of the
curvature is pleasingly mild.~\cite{Roman-ring} It is these large
metric fluctuations and associated Planck scale curvature fluctuations
that tell us that we should no longer trust semi-classical quantum
gravity behind the reliability boundary.~\cite{Reliable}

{\bf Definition 2:} An improvement of the previous definition, if the
manifold in question is multiply connected one, is to keep track of
the winding number of the geodesic. (In spacetimes containing
traversable wormholes~\cite{Morris-Thorne}$^{\!,\,}$\cite{Visser} this
will just be the number of times the geodesic threads through one of
the wormholes.) Decompose the homotopy classes of self-intersecting
geodesics emanating from the point $x$ into equivalence classes
$\Gamma_N$ characterized by winding number $N$, and define
\begin{equation}
\Omega_N(\ell^2) \equiv 
\left\{x:\exists \gamma\in \Gamma_N| 
\sigma_\gamma(x,x) \leq N^2\ell^2\right\}.
\end{equation}

\noindent
The point is that if a geodesic wraps through $N$ wormholes and is
of length less than $N\ell$, then at least one
wormhole-to-wormhole segment of the curve must be of length less
than $\ell$. Now simply replace $\Omega(\ell^2)$ by
\begin{equation}
\Omega_\infty(\ell^2) \equiv 
\Omega(\ell^2) \cup (\cup_{N=1}^\infty \Omega_N(\ell^2))
\end{equation}

\noindent
in all definitions regarding the reliability region.

\medskip

{\bf Definition 3:} The definition given above still does not
capture all of the situations in which we should cease trusting
semi-classical quantum gravity.  We should also not trust regions
where the background manifold exhibits Planck scale curvature. We
might naively decide to look at sets such as

\begin{equation}
\Omega_R(\ell^2) = 
\left\{x: R_{\mu\nu\sigma\rho} R^{\mu\nu\sigma\rho} > \ell^{-2} \right\}.
\end{equation}

\noindent
We now augment the unreliable region by setting

\begin{equation}
{\cal U} \equiv  \Omega_\infty(+\ell_{Planck}^2) 
\cup \Omega_R(+\ell_{Planck}^2).
\end{equation}

\noindent
Similarly, the reliability boundary becomes
\begin{eqnarray}
{\cal B}_{Planck} &\equiv& \partial[{\cal U}]
\; \equiv \;
\partial \left[ \Omega_\infty(+\ell_{Planck}^2)
\cup \Omega_R(+\ell_{Planck}^2) \right],
\end{eqnarray}

\noindent
and the reliability horizon becomes
\begin{eqnarray}
{\cal H}^+_{Planck} &\equiv& \partial[J^+({\cal U})]
\; \equiv \;
\partial[J^+\left(\Omega_\infty(+\ell_{Planck}^2) 
\cup \Omega_R(+\ell_{Planck}^2) \right)].
\end{eqnarray}

\section{Discussion}

The reliability boundary has important applications to Hawking's
chronology protection conjecture,~\cite{Hawking:cpc,Kim-Thorne,Hawking:dos}
and to the related Kay--Radzikowski--Wald singularity
theorems.~\cite{KRW,Cramer-Kay} I argue~\cite{Reliable} that the
chronology horizon is always hidden by the reliability horizon and
that semiclassical quantum gravity should fail before reaching the
chronology horizon.

\section*{Acknowledgements}

This research was supported by the U.S. Department of Energy.

\section*{References}

\end{document}